\documentclass[aps,prd,showpacs,nofootinbib,preprintnumbers,onecolumn]{revtex4}

\usepackage{bm}
\usepackage{latexsym}
\usepackage{dcolumn}
\usepackage{amsfonts,amssymb,amsmath}
\usepackage{graphicx,epsfig}
\usepackage{psfrag}
\usepackage{multirow}

\def\beq{\begin{equation}}
\def\eeq{\end{equation}}
\def\br{\begin{eqnarray}}
\def\er{\end{eqnarray}}
\def\benu{\begin{enumerate}}
\def\eenu{\end{enumerate}}

\def\l{\left}
\def\r{\right}
\def\d{{\rm d}}
\def\eq#1{{Eq.~(\ref{#1})}}

\begin{document}
\title[The tensor-to-scalar ratio in punctuated inflation]{The
tensor-to-scalar ratio in punctuated inflation}
\author{Rajeev Kumar Jain$^{1}$\footnote{Current address:~Department of
Theoretical Physics, University of Geneva, 24~Quai Ernest-Ansermet,
CH-1211, Geneva 4, Switzerland. E-mail:~rajeev.jain@unige.ch}}
\author{Pravabati Chingangbam$^{2}$\footnote{Current address:~Astrophysical 
Research Center for the Structure and Evolution of the Cosmos, Sejong 
University, 98 Gunja Dong, Gwangjin gu, Seoul 143-747, Korea.
E-mail:~prava@kias.re.kr}}
\author{L.~Sriramkumar$^{1}$\footnote{E-mail:~sriram@hri.res.in}}
\author{Tarun Souradeep$^{3}$\footnote{E-mail:~tarun@iucaa.ernet.in}}
\affiliation{$^{1}$Harish-Chandra Research Institute, Chhatnag Road,
Jhunsi, Allahabad~211~019, India.\\
$^{2}$Korea Institute for Advanced Study, 207--43 Cheongnyangni 2-dong,
${}$~Dongdaemun-gu, Seoul 130-722, Korea.\\
$^{3}$IUCAA, Post Bag 4, Ganeshkhind, Pune 411 007, India.}
\begin{abstract}
Recently, we have shown that scalar spectra with lower power on large
scales and certain other features naturally occur in {\it punctuated
inflation}\/, i.e. the scenario wherein a brief period of rapid roll
is sandwiched between two stages of slow roll inflation.
Such spectra gain importance due to the fact that they can lead to
a better fit of the observed Cosmic Microwave Background (CMB)
anisotropies, when compared to the conventional, featureless, power
law spectrum.
In this paper, with examples from the canonical scalar field as well
as the tachyonic models, we illustrate that, in punctuated inflation,
a drop in the scalar power on large scales is {\it always}\/
accompanied by a rise in the tensor power and, hence, an even more
pronounced increase in the tensor-to-scalar ratio~$r$ on these scales.
Interestingly, we find that~$r$ actually {\it exceeds well beyond unity}\/
over a small range of scales.
{\it To our knowledge, this work presents for the first time, examples
of single scalar field inflationary models wherein $r\gg 1$.}\/
This feature opens up interesting possibilities.
For instance, we show that the rise in $r$ on large scales translates
to a rapid increase in the angular power spectrum, $C_{\ell}^{\rm BB}$,
of the B-mode polarization of the CMB at the low multipoles.
We discuss the observational implications of these results.
\end{abstract}
\maketitle


\section{Introduction and motivation}

The concordant cosmological model---viz. a spatially flat, $\Lambda$CDM
model and a nearly scale invariant primordial spectrum, with or without
a small tensor contribution (say, with a tensor-to-scalar ratio $r$ of
less than $0.1$)---seems to fit the recent Cosmic Microwave Background
(CMB) data rather well~\cite{cmb-ro}.
However, different observations have indicated that a few low multipoles
of the observed CMB angular power spectrum lie outside the cosmic
variance associated with the concordant model~\cite{cmb-eo}.
These discrepancies have remained in subsequent updates of the
data~\cite{cmb-ro,cmb-eo}, and have also survived in other
independent estimates of the angular power spectrum (see, for
instance, Refs.~\cite{saha-2006-2008}).
Given the CMB observations, a handful of model independent approaches
have been constructed over the last few years to recover the primordial
power spectrum~\cite{rc}.
At the smaller scales, all these approaches arrive at a spectrum that
is nearly scale invariant.
However, many of the approaches seem to unambiguously point to a sharp
drop in power (with specific features) at the scales corresponding to
the Hubble scale today.

Even as the debate about the statistical significance of the outliers
in the CMB data has continued~\cite{cmbo-ss}, a considerable amount of
effort has been devoted to understand the possible physical reasons
behind these outliers (for an inexhaustive list, see
Refs.~\cite{other-p,quadrupole,quadrupole-sic,general}).
Within the inflationary paradigm, different models have been
constructed to produce a sharp drop in the scalar power at
large scales, so as to lead to a better fit to the low quadrupole
(see Refs.~\cite{quadrupole,quadrupole-sic}; for earlier efforts
that discuss generating features in the inflationary perturbation
spectrum, see Refs.~\cite{hodges-1990,starobinsky-1992,early-fips}).
However, many of the scenarios that have been considered in this
context seem rather artificial---they either assume a specific
pre-inflationary regime or specific initial conditions for the
inflaton~\cite{quadrupole-sic}.
Also, in some cases, either certain special initial conditions
are chosen for the perturbations or the initial conditions are
imposed when a subset of the modes are outside the Hubble
radius~\cite{quadrupole-sic}.
Such requirements clearly contradict the spirit of inflation.

Motivated by the aim of arriving at the desired power spectrum
without any special initial conditions on either the background or
the perturbations, we have recently considered a setting involving
two stages of slow roll inflation that sandwich an intermediate
period of departure from inflation~\cite{pi}.
In such a {\it punctuated inflationary scenario}\footnote{For the
earliest discussions on such a possibility, see
Refs.~\cite{hodges-1990,starobinsky-1992}.},\/ the first phase of
slow roll inflation allows us to impose the standard, sub-Hubble
initial conditions on the perturbations which may leave the Hubble
radius during the subsequent rapid roll regime (i.e. a period wherein
the first slow roll parameter $\epsilon \gtrapprox 1$).
The second slow roll phase lasts for, say, $50$--$60$ $e$-folds,
thereby enabling us to overcome the well known horizon problem
associated with the hot big bang model.
We had discovered that such a background behavior can be achieved
in certain large field inflationary models wherein the potentials
contain a point of inflection [the form of the potentials we had
considered are encountered in the Minimal Supersymmetric Standard
Model (MSSM)].
We had shown that the slow-rapid-slow roll transition leads to a step
like feature in the scalar power spectrum.
Importantly, we had found that, if we set the scales such that the
drop in the power spectrum occurs at a length scale that roughly
corresponds to the Hubble radius today, then a spectrum we had
obtained leads to a much better fit to the WMAP $5$-year data when
compared to the best fit reference $\Lambda$CDM model with the
standard, power law, primordial spectrum~\cite{pi}.

All models of inflation generate tensor perturbations that can
potentially have an observable effect on the measured CMB
temperature and polarization spectra~\cite{tensors}.
Barring an exception~\cite{nicholson-2008}, most of the efforts in
the literature have focused on suppressing the scalar power spectrum
on large scales, and have overlooked the corresponding effects on the
tensors.
In this paper, we investigate the effects of the slow-rapid-slow
roll transition on the tensor perturbations in the canonical scalar
field and the tachyonic~\cite{tachyon,sami-prava,steer-2004}
inflationary models.
Aided by a few different examples (including the specific model that
we had considered earlier), we show that, in punctuated inflation,
a drop in the scalar power on large scales is {\it always}\/ associated
with an increase in the tensor power and, hence, a dramatic rise in the
tensor-to-scalar ratio $r$, on these scales.
In fact, we find that the strong rise leads to a small range of modes
for which the tensor-to-scalar ratio actually proves to be {\it much
greater than unity}\footnote{In the models we consider, $r$ attains a
maximum value of about $100$.
Though the tensor-to-scalar ratio is large, the actual amplitude of
the tensor perturbations still remains small enough for the linear
perturbation theory to be valid.}.\/
{\it We believe that this is the first instance in the literature
wherein examples of single scalar field inflationary models resulting
in $r\gg1$ are being presented.}\/
However, if we are to utilize the drop in the scalar power to provide
a better fit to the low CMB quadrupole, then the modes with the rather
large tensor-to-scalar ratio turn out to be bigger than the Hubble
scale today.

The rapid rise in the tensor-to-scalar ratio~$r$ at large scales
translates to a dramatic enhancement in the angular power spectrum,
$C_{\ell}^{\rm BB}$, of the B-mode polarization of the CMB at the
low multipoles.
This could potentially be a characteristic signature of punctuated
inflationary scenarios that match the CMB data well.
But, in the specific models of punctuated inflation that we have
explored to match the low multipoles of CMB temperature power spectrum,
the enhanced $C_{\ell}^{\rm BB}$ is not at an observable level.
This is due to the following two reasons.
Firstly, the band of scales where $r \gg 1$ is well beyond the Hubble
scale today and, secondly, because $r$ is extremely small at large
wavenumbers.
However, it is readily conceivable that there exist models of punctuated
inflation where either one or both of these features can be modified
favorably to arrive at observable levels of $C_{\ell}^{\rm BB}$.
We defer a systematic hunt for such models to a later publication
and, in this work, we highlight the extremely large values of
tensor-to-scalar ratio~$r$ attainable in the punctuated inflationary
scenario.

This paper is organized as follows.
In Sec.~\ref{sec:dsri}, after rapidly summarizing the essential
equations and quantities, we outline the broad features of the
scalar and tensor spectra in punctuated inflation.
In Sec.~\ref{sec:csf}, we discuss the spectra that arise
in two different punctuated inflationary models involving the
canonical scalar field, while, in Sec.~\ref{sec:tf}, we
discuss the spectra in a particular tachyonic model.
In Sec.~\ref{sec:clbb}, we consider the corresponding
effects on the angular power spectrum of the B-mode polarization
of the CMB.
Finally, in Sec.~\ref{sec:concl}, we conclude with a brief
discussion on the implications of this feature.
In the appendix, to highlight the feature that the tensor-to-scalar
ratio can turn out to be greater than unity for a range of modes in
punctuated inflation, we illustrate the evolution of the scalar and
tensor amplitudes for a particular mode from this domain.

In the discussions below, we shall set $\hbar$ and $c$ as well
as $M_{_{\rm P}} = (8\,\pi\, G)^{-1/2}$ to unity.
As is often done in the context of inflation, we shall work with
the spatially flat Friedmann model.
Also, throughout, an overdot and an overprime shall denote
differentiation with respect to the cosmic and the conformal
times, respectively.
Moreover, $\phi$ shall denote the scalar field described by the
canonical action, while $T$ shall denote the tachyon.


\section{Characteristics of the perturbation spectra in punctuated
inflation}\label{sec:dsri}

In this section, after outlining the equations governing the
perturbations and listing the observable quantities of interest,
we discuss the broad features of the scalar and the tensor spectra
that arise in the punctuated inflationary scenario.


\subsection{Key equations and quantities}

We begin by briefly summarizing the essential equations and the
quantities that we shall be interested in~\cite{texts,reviews}.
The curvature perturbation ${\cal R}_{k}$ and the tensor
perturbation~${\cal U}_{k}$ satisfy the differential equations
\beq
{\cal R}_{k}''+2\, \l(\frac{z'}{z}\r)\, {\cal R}_{k}'
+k^{2}\, c_{_{\rm S}}^2\, {\cal R}_k=0\quad{\rm and}\quad
{\cal U}_{k}''+2\, \l(\frac{a'}{a}\r)\, {\cal U}_{k}'
+k^{2}\; {\cal U}_k=0,\label{eq:egp}
\eeq
where $a$ is the scale factor and $c_{_{\rm S}}$ denotes the speed of
propagation of the scalar perturbations.
The effective speed of sound $c_{_{\rm S}}$ turns out to be unity for
the canonical scalar field, while $c_{_{\rm S}}^{2}=(1-{\dot T}^2)$ in
the case of the tachyon~\cite{steer-2004}.
Also, the quantity $z$ is given by
\beq
z = \l(a\, {\dot \phi}/H\r)\quad{\rm and}\quad
z=\l(\sqrt{3}\,a\, \dot{T}/c_{_{\rm S}}\r),
\eeq
in the case of the conventional scalar field and the tachyonic
inflationary models, respectively, with $H$, as usual, being
the Hubble parameter.
The scalar and the tensor power spectra ${\cal P}_{_{\rm S}}(k)$
and ${\cal P}_{_{\rm T}}(k)$ are then defined as
\beq
{\cal P}_{_{\rm S}}(k) = \l(\frac{k^{3}}{2\, \pi^{2}}\r)\,
\vert{\cal R}_{\rm k}\vert^{2}\qquad{\rm and}\qquad
{\cal P}_{_{\rm T}}(k)
= 2\, \l(\frac{k^{3}}{2\pi^{2}}\r)\,
\vert{\cal U}_{\rm k}\vert^{2},
\eeq
with the amplitude of the perturbations ${\cal R}_{k}$ and ${\cal U}_{k}$
evaluated, in general, at super-Hubble scales.
(The factor of two in the tensor spectrum ${\cal P}_{_{\rm T}}(k)$
above is to account for the two states of polarization of the
gravitational waves.)
Finally, the tensor-to-scalar ratio $r(k)$ is defined as follows:
\beq
r(k) \equiv \l(\frac{{\cal P}_{_{\rm T}}(k)} {{\cal P}_{_{\rm S}}(k)}\r).
\eeq


\subsection{The scalar and the tensor spectra in punctuated
inflation}

While considering single scalar field models, it is often remarked
that, during inflation, the amplitude of the curvature perturbations
freezes at its value at Hubble exit.
Actually, this happens to be true only if there is no departure
from slow roll inflation soon after the modes leave the Hubble
radius~\cite{she}.
But, when there is a period of deviation from slow roll, then,
it is found that the asymptotic (i.e. the extreme super-Hubble)
amplitude of the modes that exit the Hubble scale just before
the deviation are enhanced when compared to their value at Hubble
exit.
While modes that leave well before the departure from slow roll are
unaffected, it has been shown that there exists an intermediate range
of modes whose amplitudes are suppressed at super-Hubble scales.
Due to these behavior, punctuated inflation leads to a step like
feature in the scalar power spectrum.
Evidently, the two nearly flat regions of the step correspond to
modes that exit the Hubble scale during the two stages of slow roll.
For instance, in the case of the canonical scalar field models, these 
slow roll amplitudes will be given by the following standard expression
(see, for example, Refs.~\cite{texts,reviews}):
\beq
{\cal P}_{_{\rm S}}(k)
\simeq \l(\frac{1}{12\,\pi^2}\r)\,
\l(\frac{V^3}{V_\phi^2}\r)_{k=(a\,H)},\label{eq:srsa}
\eeq
where $V_\phi\equiv({\rm d}V/{\rm d}\phi)$, and the spectral amplitude 
has to be evaluated when the modes leave the Hubble radius.
The step actually contains a sharp dip before the rise, and this
feature is associated with the modes that leave the Hubble radius
just before the transition to the rapid roll regime.

Let us now understand the tensor spectrum that can result in a
similar situation.
In the case of the scalar modes, the quantity $(z'/z)$ that
appears in the differential equation for the curvature
perturbation ${\cal R}_{k}$ in~\eq{eq:egp} turns out to be
negative during a period of fast roll, and it is this feature
that proves to be responsible for the amplification or the
suppression of the modes at super-Hubble scales~\cite{she}.
In contrast, the coefficient of the friction term in the
equation for the tensor amplitude ${\cal U}_{k}$
in \eq{eq:egp}---viz. $(a'/a)$---is a positive definite
quantity at all times.
Hence, we do not expect any non-trivial super-Hubble evolution
of the tensor perturbations.
However, recall that, during a period of slow roll, the tensor 
amplitude is proportional to the potential of the scalar field 
and, in the case of the canonical scalar field models, it is 
given by~\cite{texts,reviews}
\beq
{\cal P}_{_{\rm T}}(k)
\simeq \l(\frac{2\, V}{3\, \pi^2}\r)_{k=(a\,H)}.\label{eq:srta}
\eeq
It is then immediately clear that, in the slow-rapid-slow roll
scenario of our interest, the tensor spectrum will also be in
the shape of a step, with the modes that leave during the second
slow roll phase having lower power than those which exit during
the first phase [since, unless the potential is negative, the
inflaton always rolls {\it down}\/ the potential (see, for
example, Ref.~\cite{roll-down})].
In other words, in punctuated inflation, the tensor step happens
to be in exactly the opposite direction as the step in the scalar
spectrum.
The fact that the scalar power drops at large scales, while the
tensor power rises on these scales, leads to a sharp increase
in the tensor-to-scalar ratio~$r$.
Interestingly, we find that the steep rise can result in the
tensor-to-scalar ratio being greater than unity for a small range
of modes.
(These range of modes correspond to those for which the scalar
spectrum exhibits a sharp dip before the rise.)
However, as we shall discuss below, in the specific models of
punctuated inflation that we consider, in spite of the rise, the
tensor-to-scalar ratio remains too small to be observed ($r$
proves to be less than $10^{-4}$) for the modes of cosmological
interest (say, $10^{-4} < k < 1\; {\rm Mpc}^{-1}$).
But, we believe that the increase in the tensor-to-scalar ratio at
large scales considerably improves the prospects of constructing
punctuated inflationary models wherein $C_{\ell}^{\rm BB}$ at
the low multipoles is within the observational reach of current
missions such as PLANCK~\cite{planck} or future ones such as, for
instance, CMBPol~\cite{CMBPol}.

In the following two sections, we shall explicitly illustrate
these behavior with the help of specific examples.


\section{Punctuated inflation with canonical scalar
fields}\label{sec:csf}

In this section, we shall discuss punctuated inflationary scenarios
in models where inflation is driven by the canonical scalar field.
We shall first present the model that we had considered
earlier~\cite{pi}, and then discuss a hybrid inflation model.

Before proceeding to discuss the specific models, we shall outline
as to how one can arrive at the potential and the parameters that
result in punctuated inflation and the desired scalar spectrum.
Needless to say, not all potentials will allow punctuated inflation.
Therefore, to begin with, one has to identify a potential,
or a class of potentials, that lead to such a scenario.
Even amongst the limited class of potentials, the required
slow-rapid-slow roll transition may occur only for a certain range
of values of the parameters describing the potential.
The form of the potential and the range of the parameters can
be arrived at, say, based on the behavior of the first two
potential slow roll parameters.
Once the potential and the range of the parameters that allow
punctuated inflation have been identified, we need to ensure
that the following two observational requirements are also
satisfied.
Firstly, the second stage of slow roll inflation has to last for
about $60$ $e$-folds in order to overcome the horizon problem.
Secondly, the nearly scale invariant higher step in the scalar
power spectrum has to match the COBE amplitude.
These two conditions further restrict the allowed range of the
parameters describing the potential.


\subsection{The model motivated by MSSM}

The model motivated by MSSM that we had considered in our earlier
work contains two parameters $m$ and $\lambda$, and is described
by the potential~\cite{mssm}
\beq
V(\phi) = \l(\frac{m^2}{2}\r)\,\phi^2
- \l(\frac{\sqrt{2\,\lambda\,(n-1)}\,m}{n}\r)\, \phi^n
+ \l(\frac{\lambda}{4}\r)\,\phi^{2(n-1)},
\label{eq:mssm-p}
\eeq
where $n>2$ is an integer.
This potential has a point of inflection at $\phi = \phi_0$
(i.e. the location where both $V_{\phi}$
and $V_{\phi\phi}\equiv (\d^{2}V/\d\phi^{2})$ vanish), with
$\phi_{0}$ given by
\beq
\phi_{0}
= \left[\frac{2\,m^2}{(n-1)\,\lambda}\right]^{\frac{1}{2\,(n-2)}}.
\eeq
Note that the potential~(\ref{eq:mssm-p}) reduces to a typical
large field model when the field is sufficiently far away from
the point of inflection.
It is then clear that the first stage of slow roll can be achieved
in the domain $\phi\gg \phi_{0}$, and a period of rapid roll can
occur when $\phi \simeq \l[\sqrt{2}\; (n-1)\r]$.
Also, since the first two potential slow roll parameters vanish at
the point of inflection, a second stage of slow roll can be expected
to arise when the field is very close to $\phi_0$.
We find that restarting inflation after the rapid roll phase and
the number of $e$-folds that can be achieved during the second
stage of slow roll crucially depends on the location of the point
of inflection.
We depend on the numerics to arrive at a suitable value of $\phi_0$.
Once $\phi_0$ has thus been identified, we find that the COBE
normalization determines the value of the other free parameter~$m$.

\begin{figure}[!htb]
\begin{center}
\vskip 15 pt
\resizebox{210pt}{150pt}{\includegraphics{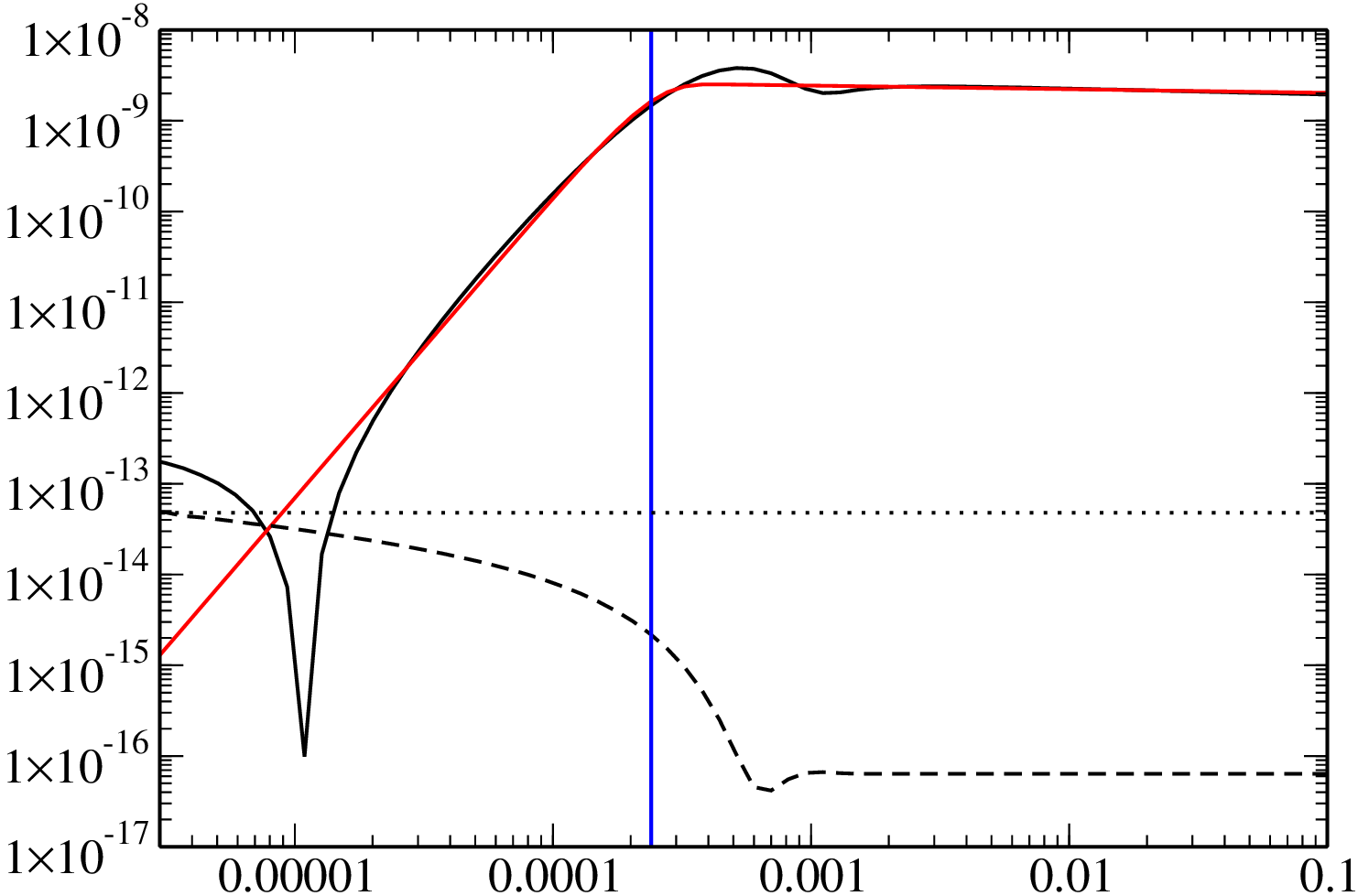}}
\hskip 20pt
\resizebox{210pt}{150pt}{\includegraphics{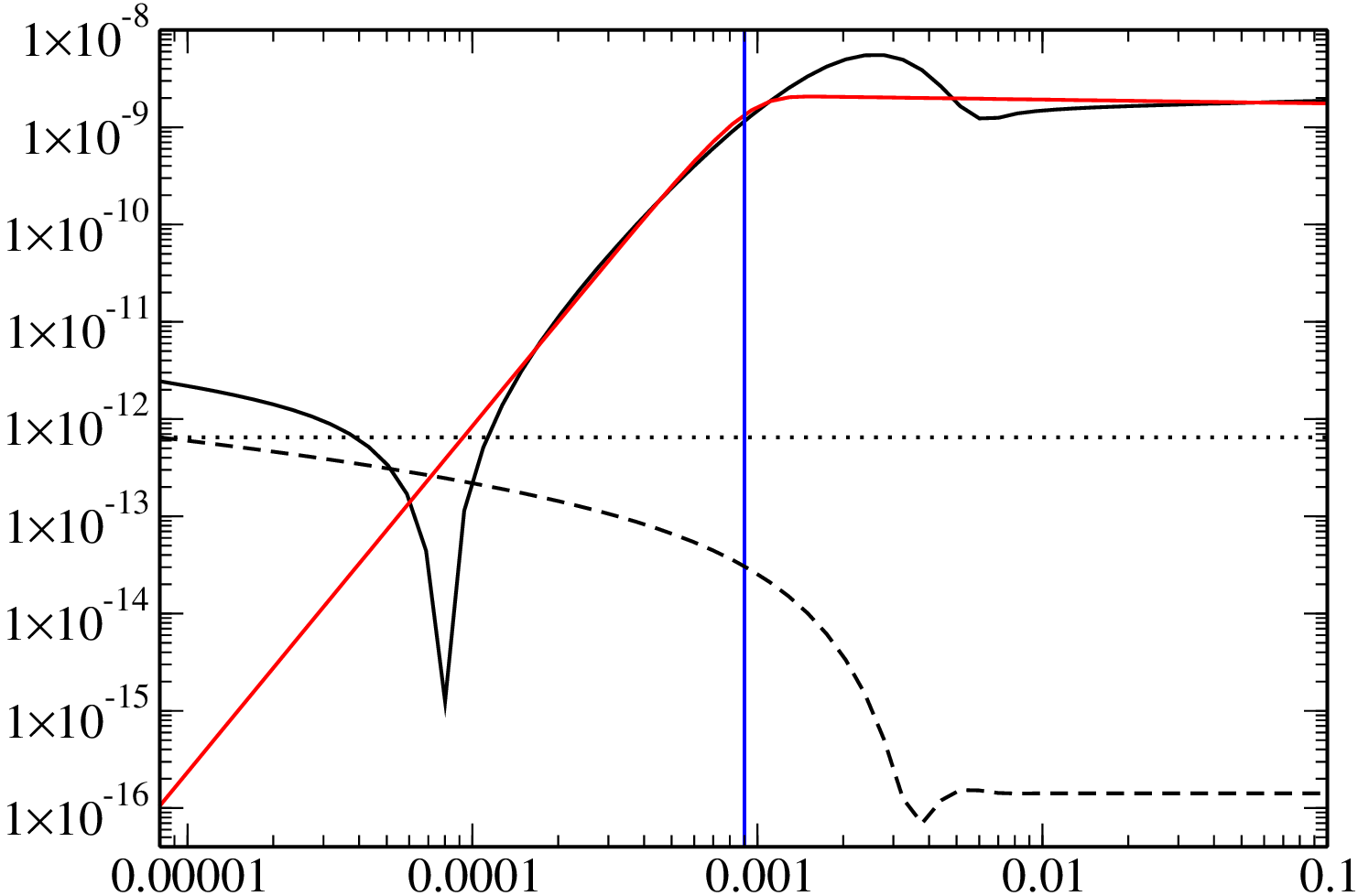}}
\vskip -135 true pt \hskip -232 true pt
\rotatebox{90}{${\cal P}_{_{\rm S}}(k)$, ${\cal P}_{_{\rm T}}(k)$}
\hskip 218 true pt
\rotatebox{90}{${\cal P}_{_{\rm S}}(k)$, ${\cal P}_{_{\rm T}}(k)$}
\vskip 85 true pt
\hskip 10 true pt $k$ \hskip 230 true pt $k$
\caption{The scalar power spectrum ${\cal P}_{_{\rm S}}(k)$
(the solid black line) and the tensor power spectrum
${\cal P}_{_{\rm T}}(k)$ (the dashed black line) have been
plotted as a function of the wavenumber~$k$ for the cases of
$n=3$ (on the left) and $n=4$ (on the right).
These spectra correspond to the following values of the potential 
parameters: $m=1.5368\times10^{-7}$ and $\lambda=6.1517\times10^{-15}$
(corresponding to $\phi_{0}=1.9594$) for $n=3$ case and 
$m=1.1406\times10^{-7}$ and $\lambda=1.448\times10^{-16}$
(corresponding to $\phi_{0}=2.7818$) for $n=4$ case.
The red curve in these plots is the spectrum~(\ref{eq:ps-ecm})
with the exponential cut off, whose parameters have been arrived
at by a simple visual comparison with the numerically evaluated
scalar spectrum.
It corresponds to $A_{_{\rm S}}= 2\times10^{-9}$, $n_{_{\rm S}}= 0.945$, 
$\alpha=3.35$ and $k_{\ast} = 2.4\times 10^{-4}\;{\rm Mpc}^{-1}$ in the 
$n=3$ case, while $A_{_{\rm S}}= 2\times10^{-9}$, $n_{_{\rm S}}= 0.95$, 
$\alpha=3.6$ and $k_{\ast} = 9.0\times 10^{-4}\;{\rm Mpc}^{-1}$ in the 
case of $n=4$.
Note that the vertical blue lines denote $k_{\ast}$.
We should mention that, in the two slow roll regimes, the spectral 
amplitudes evaluated in the slow roll approximation 
[cf. Eqs.~(\ref{eq:srsa}) and~(\ref{eq:srta})] match the above exact 
numerical spectra quite well. 
The horizontal dotted lines indicate the maximum value of the tensor 
amplitude that can arise in these MSSM-motivated, punctuated
inflationary models.}\label{fig:ps-n34}
\end{center}
\end{figure}

In Fig.~\ref{fig:ps-n34}, we have plotted the scalar and the tensor
power spectra for the cases of $n=3$ and $n=4$.
These spectra correspond to the parameters that provide the best fit
to the WMAP $5$-year data (for further details, see Ref.~\cite{pi}).
We should mention that in these cases inflation is actually
interrupted for about one $e$-fold during the rapid roll regime.
We had found that, while the $n=3$ case provides a much better fit
to the data than the reference concordant model, the $n=4$ case
leads to a very poor fit to the data.
We believe that the poor fit by the $n=4$ case can be attributed
to the large bump in the scalar power spectrum that arises just
before it turns nearly scale invariant.
Since the bump grows with $n$, we feel that the cases with $n > 4$
will fit the data much more poorly and, hence, we have not compared
these cases with the data.

The scalar power spectrum with a drop in power at large scales is
often approximated by an expression with an exponential cut off of
the following form (see, for instance, the first two references in
Ref.~\cite{quadrupole-sic}):
\beq
{\cal P}_{_{\rm S}}(k)
=A_{_{\rm S}}\, \biggl(1-\exp{\l[-(k/k_{\ast})^{\alpha}\r]}\biggr)\,
k^{n_{_{\rm S}}-1}.\label{eq:ps-ecm}
\eeq
In Fig.~\ref{fig:ps-n34}, we have also plotted this expression for
values of $A_{_{\rm S}}$, $n_{_{\rm S}}$, $\alpha$ and $k_{\ast}$
that closely approximate the exact spectra we obtain.

\begin{figure}[!htb]
\begin{center}
\vskip 25 pt
\resizebox{270pt}{180pt}{\includegraphics{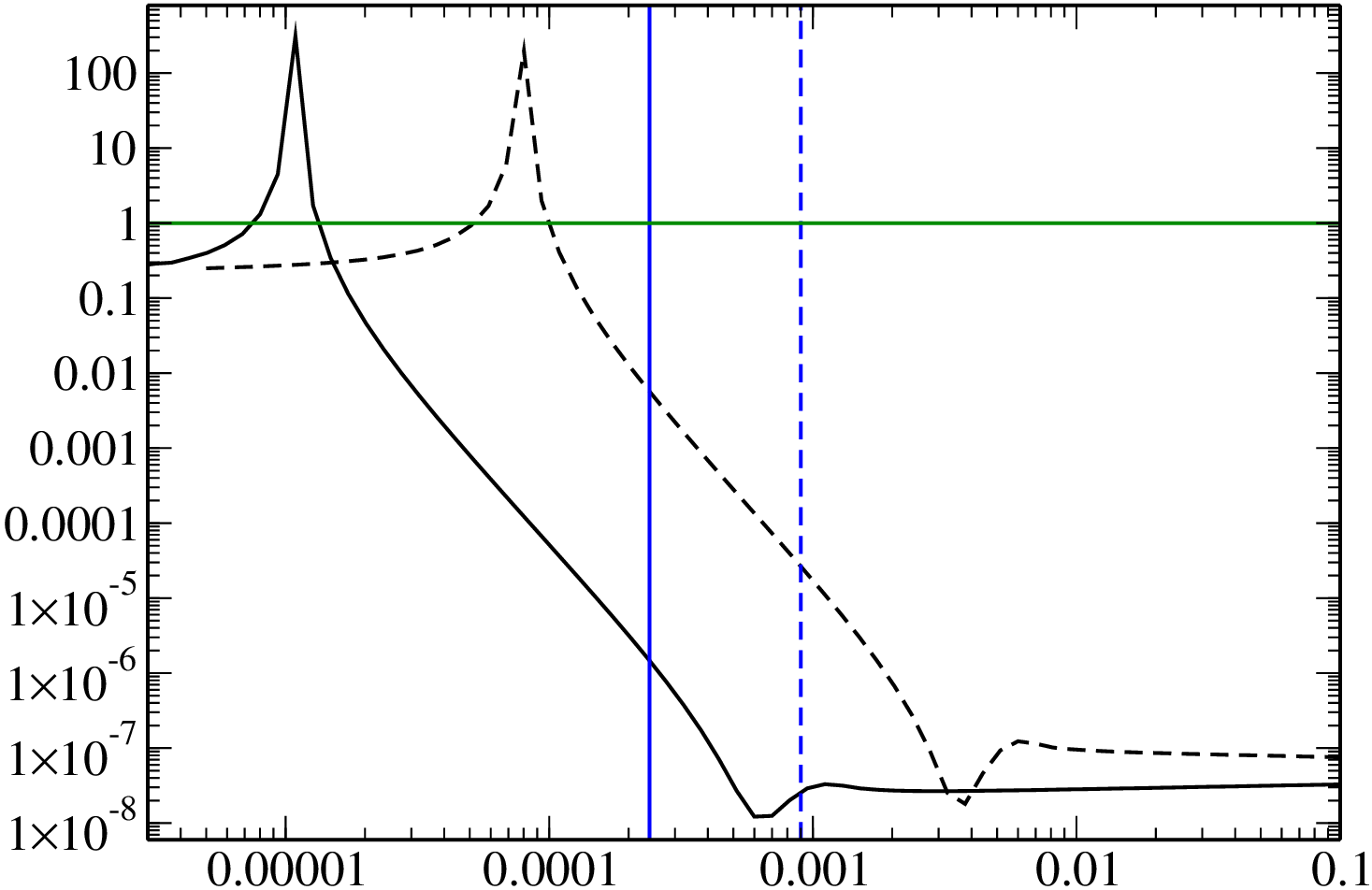}}
\vskip -145 true pt \hskip -290 true pt $r(k)$
\vskip 135 true pt \hskip 10 true pt $k$
\caption{The tensor-to-scalar ratio $r(k)$ for the cases of $n=3$
(the solid black line) and $n=4$ (the dashed black line) has been
plotted as a function of the wavenumber~$k$.
These plots have been drawn for the same choice of parameters
as in the previous figure.
The vertical solid and dashed blue lines denote the $k_{\ast}$
corresponding to the $n=3$ and $n=4$ cases, respectively.
Note that, despite the rise at the larger wavelengths, the
tensor-to-scalar ratio remains smaller than $10^{-4}$ for
modes of cosmological interest (i.e. for $k\gtrsim k_{\ast}$).
For this reason, in our earlier work~\cite{pi}, we had ignored
the tensor contribution when comparing with the WMAP $5$-year data.
Interestingly, we find that there arises a domain wherein the
tensor-to-scalar ratio $r(k)$ is actually {\it much greater
than unity}.\/
To highlight this feature, we have included the horizontal green
line which denotes $r=1$.
In the appendix, we have plotted the evolution of the scalar
and tensor amplitudes for a mode from this domain.}
\label{fig:r}
\end{center}
\end{figure}

In Fig.~\ref{fig:r}, we have plotted the resulting tensor-to-scalar
ratio~$r$ for the two cases of $n=3$ and $n=4$.
Clearly, the broad characteristics of the scalar and the tensor
spectra as well as the tensor-to-scalar ratio that we had outlined
in the previous section are corroborated by these two figures.


\subsection{A hybrid inflation model}

Another model that is known to lead to a punctuated inflationary
scenario is a hybrid model that can be effectively described by the
following potential (see the first reference in Ref.~\cite{she};
for the earliest discussion of the model, see Ref.~\cite{linde-1994}):
\beq
V(\phi) =\l(\frac{M^4}{4}\r)\, \l(1+B\, \phi^4\r).
\label{eq:hip}
\eeq
For suitable values of the parameter $B$, this potential admits two
stages of slow roll inflation, broken by a brief period of rapid roll.
The first slow roll phase is driven by the $\phi^4$ term and, when
$\phi$ has rolled down the potential and has become sufficiently
small, the false vacuum term drives the second phase.
The parameter $M$ determines the amplitude of nearly scale invariant
lower step in the scalar spectrum (associated with the modes that leave
during the first stage of slow roll inflation), with a mild dependence
on $B$.
However, $B$ very strongly affects the rise in the scalar power
(corresponding to the modes that leave just before the rapid
roll stage) and the asymptotic spectral index (associated with
the modes that leave during the second stage of inflation),
since it determines the extent and the duration of the departure
from slow roll.

\begin{figure}[!htb]
\begin{center}
\vskip 20 pt
\resizebox{210pt}{150pt}{\includegraphics{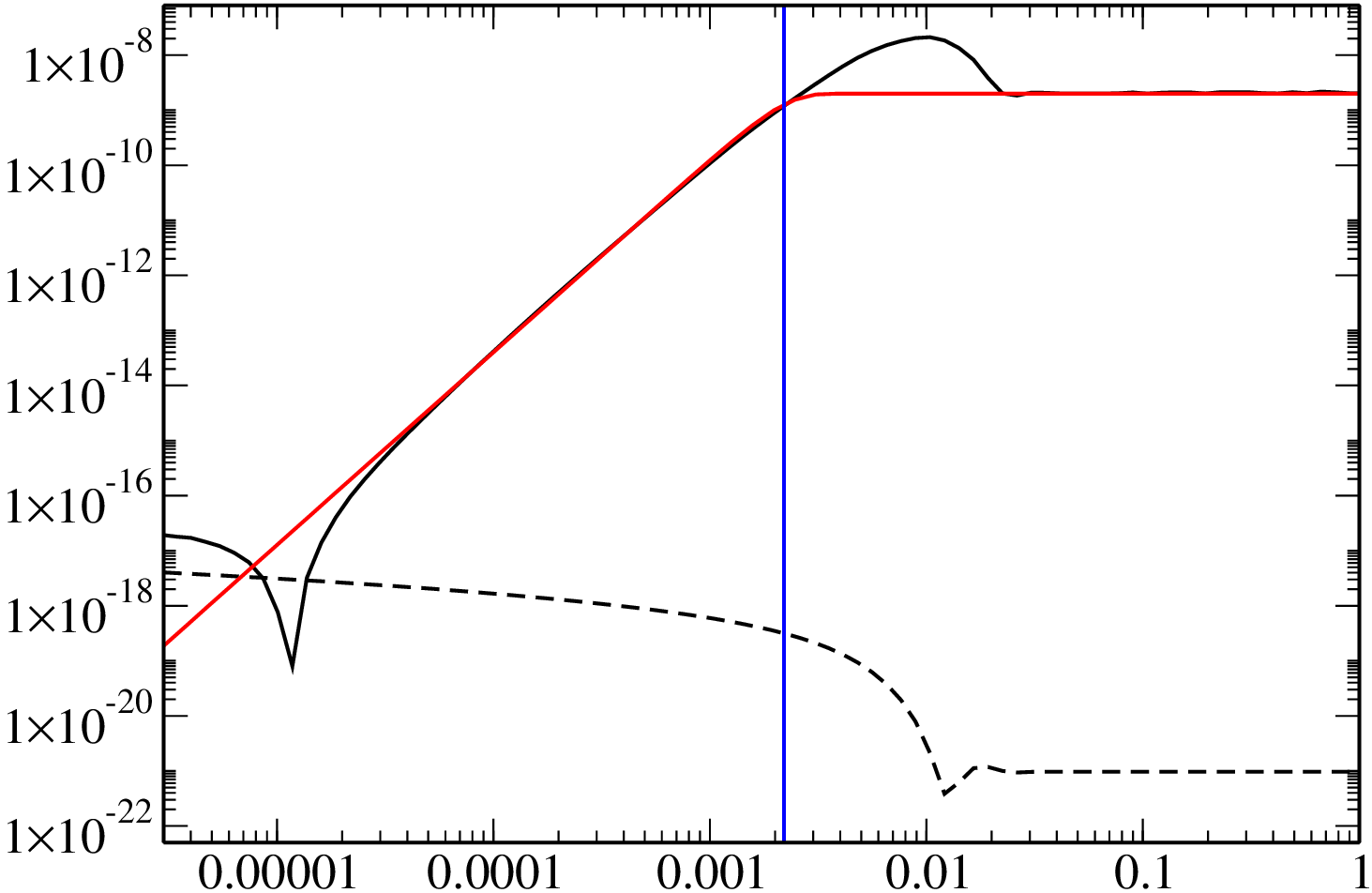}}
\hskip 20pt
\resizebox{210pt}{150pt}{\includegraphics{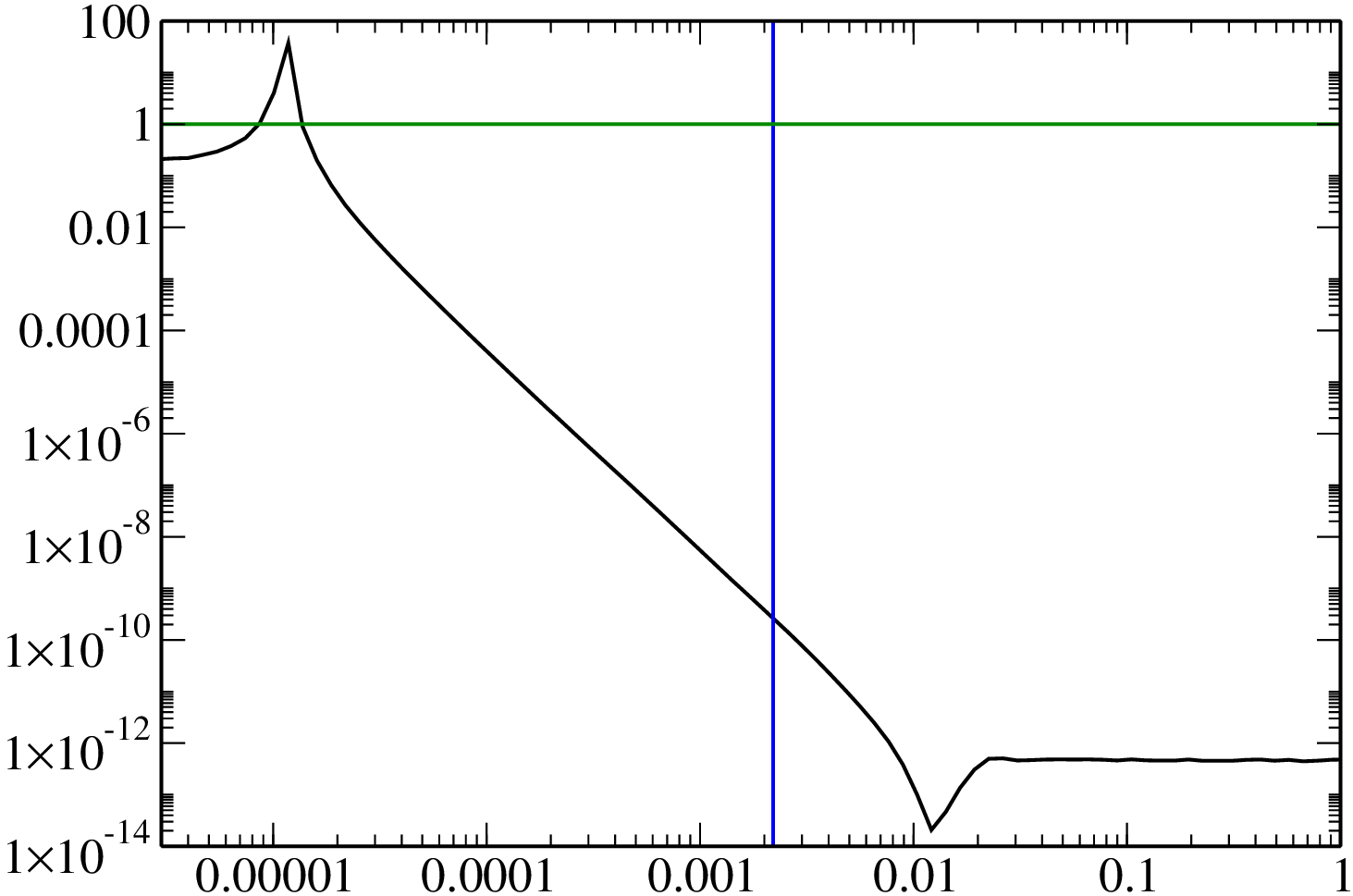}}
\vskip -142 true pt \hskip 16 true pt $r(k)$
\vskip -12 pt \hskip -465 true pt
\rotatebox{90}{${\cal P}_{_{\rm S}}(k)$, ${\cal P}_{_{\rm T}}(k)$}
\vskip 90 pt \hskip 15 true pt $k$ \hskip 235 true pt $k$
\caption{The scalar power spectrum ${\cal P}_{_{\rm S}}(k)$ (the
solid black line) and the tensor power spectrum ${\cal P}_{_{\rm T}}(k)$
(the dashed black line) have been plotted (on the left) as a function
of the wavenumber~$k$ for the hybrid inflation model described by the
potential~(\ref{eq:hip}).
The corresponding tensor-to-scalar ratio $r(k)$ has also been plotted
(on the right).
As in the previous figure, the horizontal green line in the right
graph denotes $r=1$.
These spectra correspond to following values of the potential
parameters: $M=2.6\times 10^{-5}$ and $B=0.552$.
The solid red curve in the left graph is the exponential cut off
spectrum~(\ref{eq:ps-ecm}) corresponding to $A_{_{\rm S}}= 2\times
10^{-9}$, $n_{_{\rm S}}= 1.0$, $k_{\ast} = 2.2\times 10^{-3}\;
{\rm Mpc}^{-1}$ and $\alpha=3.5$.
The vertical blue line in both the graphs denotes $k_{\ast}$.
We should mention that the blue tilt in the scalar spectrum is very
small and, hence, is not evident from the figure.
Moreover, we find that, as in the previous MSSM-motivated examples, 
in the two slow roll regimes, the amplitudes of the spectra calculated 
in the slow roll approximation agree very well with the exact numerical 
spectra.
Further, it should be noted that the amplitude of the tensors in the first 
slow roll phase determines the maximum tensor amplitude that can arise in 
such a punctuated inflationary scenario.}
\label{fig:leachpower}
\end{center}
\end{figure}

We are able to achieve COBE normalization for a suitable combination
of the parameters $M$ and $B$.
For these values of the parameters, we find that, as in the MSSM-motivated
model, a departure from inflation occurs (again, for about one
$e$-fold) during the rapid roll phase.
In Fig.~\ref{fig:leachpower}, we have plotted the resulting scalar
and the tensor power spectra as well as the associated tensor-to-scalar
ratio.
We should hasten to clarify that we have not compared the hybrid model
with the CMB data, as we had done in the $n=3$ and $n=4$ cases of the
MSSM-motivated model.
A well known property of the hybrid models is that they lead to blue
scalar spectra.
We believe that, the blue tilt, along with the rather large bump (which
turns out to be larger than the one in $n=4$, MSSM-motivated model) will 
considerably spoil the fit to the CMB data.


\section{An example of tachyonic punctuated inflation}\label{sec:tf}

In this section, we shall consider a tachyonic model that allows
punctuated inflation.
Since our experience suggests that a point of inflection in
the potential is an assured way of achieving a slow-rapid-slow
roll transition, we shall construct a tachyonic potential
containing a point of inflection.

Tachyonic potentials are usually written in terms of two
parameters, say, $\lambda$ and $T_{0}$, in the following
form~\cite{tachyon,sami-prava,steer-2004}:
\beq
V(T)=\lambda\; V_{1}(T/T_{0}),
\eeq
where $V_{1}(T/T_{0})$ is a function which has maximum at the
origin and vanishes as $T\to\infty$.
In order to achieve the necessary amount of inflation and the correct
amplitude for the scalar perturbations, suitable values for the two
parameters $\lambda$ and $T_{0}$ that describe the above potential
can be arrived at as follows.
One finds that, in these potentials, inflation typically occurs
around $T\simeq T_{0}$ corresponding to an energy scale of about
$\lambda^{1/4}$.
Moreover, it turns out that, the quantity $(\lambda\, T_{0}^2)$ has
to be much larger than unity (in units wherein $M_{_{\rm P}}=1$) for
the potential slow roll parameters to be small and, thereby ensure
that, at least, $60$ $e$-folds of inflation takes place.
One first chooses a sufficiently large value of $(\lambda\, T_{0}^2)$
by hand in order to guarantee slow roll.
The COBE normalization condition for the scalar perturbations then
provides the second constraint, thereby determining the values of
both the parameters~$\lambda$ and $T_{0}$~\cite{steer-2004}.

\begin{figure}[!htb]
\begin{center}
\vskip 15 pt
\resizebox{210pt}{150pt}{\includegraphics{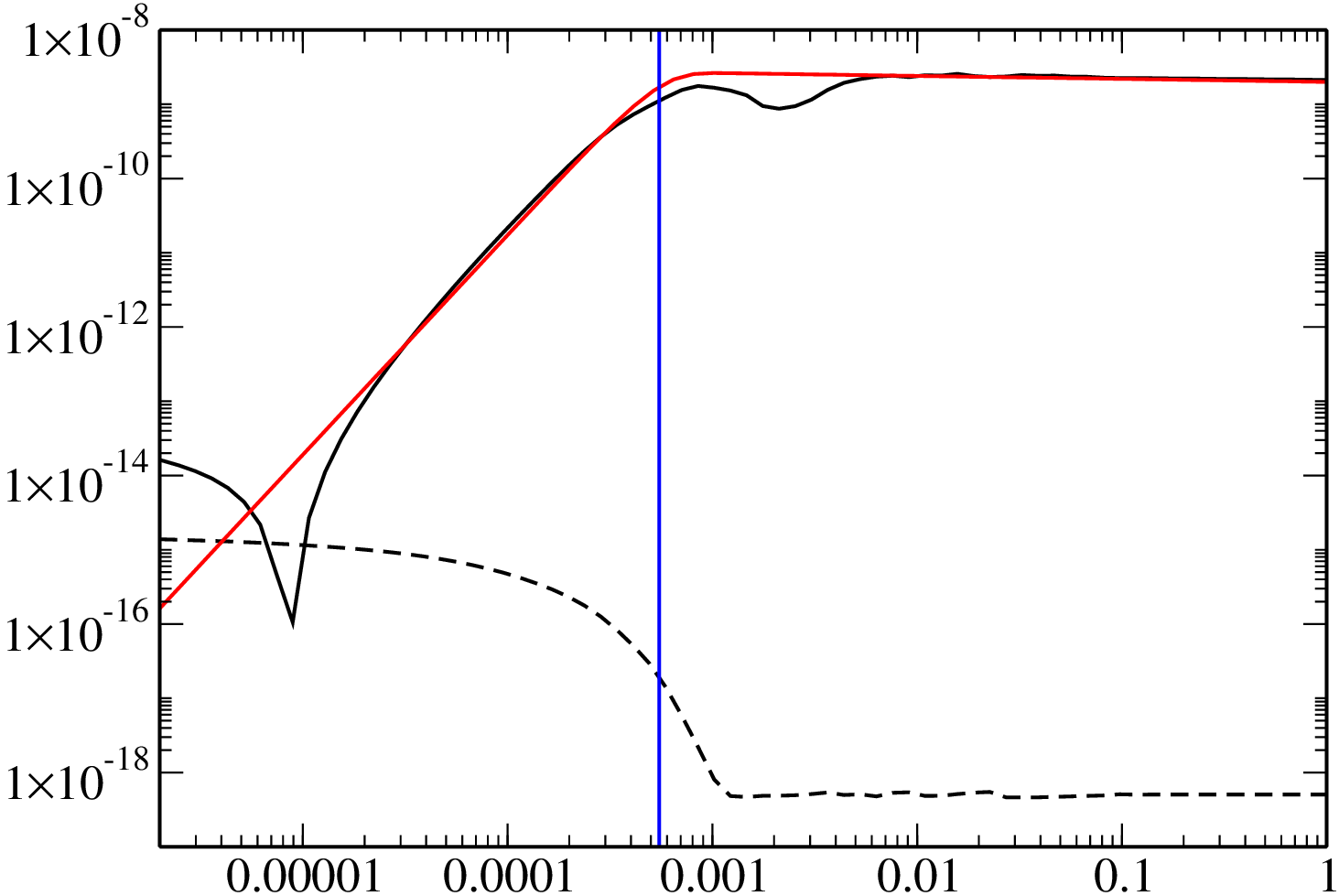}}
\hskip 20pt
\resizebox{210pt}{150pt}{\includegraphics{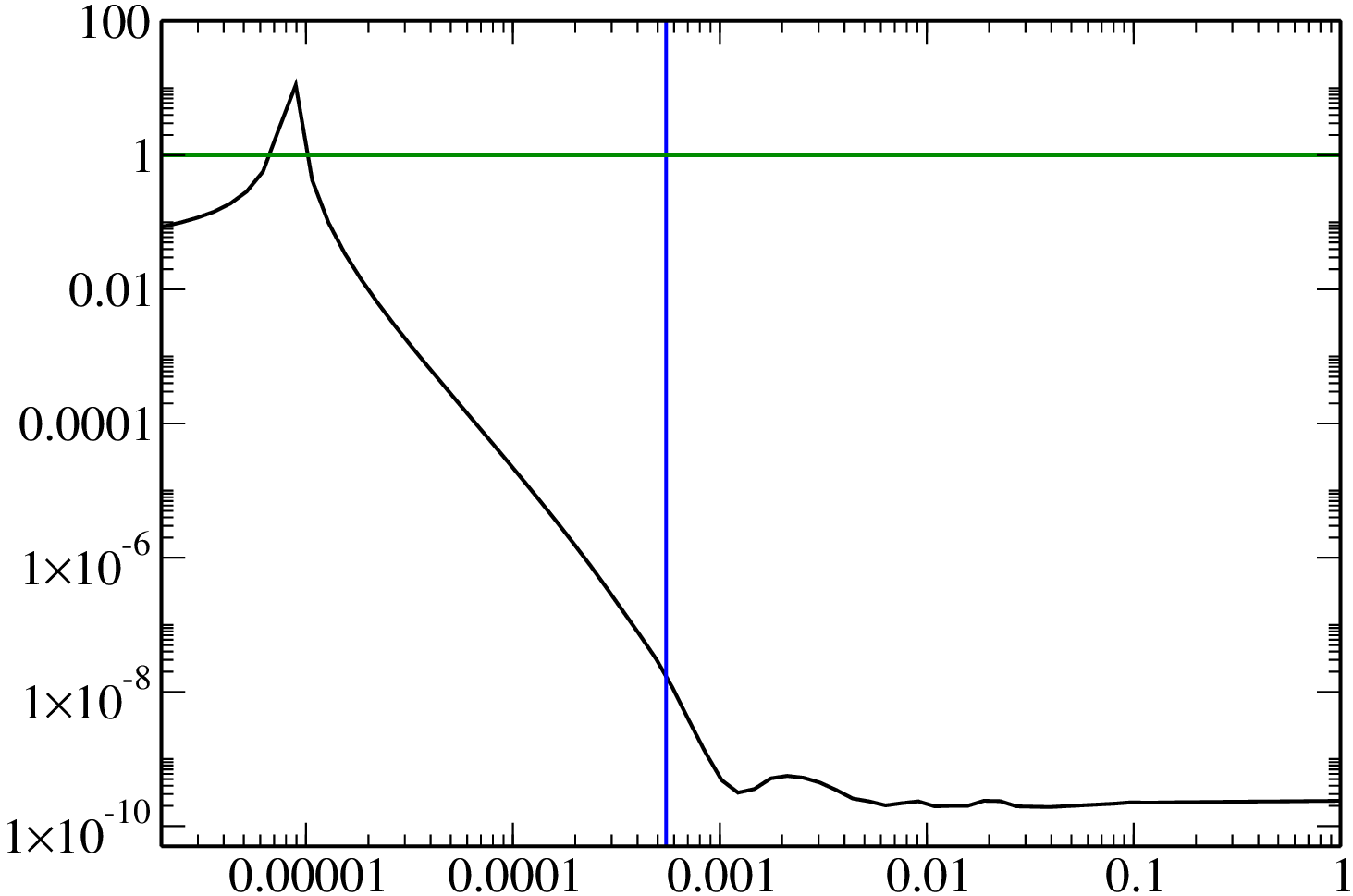}}
\vskip -142 true pt \hskip 16 true pt $r(k)$
\vskip -12 pt \hskip -465 true pt
\rotatebox{90}{${\cal P}_{_{\rm S}}(k)$, ${\cal P}_{_{\rm T}}(k)$}
\vskip 90 pt \hskip 15 true pt $k$ \hskip 235 true pt $k$
\caption{The scalar and the tensor power spectra, the corresponding
tensor-to-scalar ratio as well as the exponential cut off spectrum
have been plotted exactly in the same fashion as in the previous
figure.
These figures correspond to the following values of the parameters:
$\lambda=10^{-13}$, $T_{0}=3.55\times 10^{7}$, $x_{1}=10$,
$A_{_{\rm S}}= 2\times 10^{-9}$, $n_{_{\rm S}}= 0.96$, $k_{\ast}
= 5.5\times 10^{-4}\, {\rm Mpc}^{-1}$, and $\alpha=3$.
We should also point out that, as in the earlier cases, in the two 
slow roll regimes, the spectral amplitudes, when evaluated in the 
slow roll approximation, are in good agreement with the exact 
numerical spectra.
Again, as in the previous examples, the maximum value of the tensor
amplitude that can arise in such punctuated inflationary scenarios is
determined by its value in the first slow roll phase.}
\label{fig:ti}
\end{center}
\end{figure}

Now, consider a tachyon potential of the form
\beq
V(x) = \l(\frac{\lambda}{1+ g(x)}\r),
\eeq
where $x=(T/T_{0})$.
Let the function $g(x)$ be defined as an integral of yet another
function $f(x)$ as follows:
\beq
g(x) = \int \d x\; f(x),
\eeq
with the constant of integration assumed to be zero.
If we choose $f(x)$ to be a polynomial that vanishes at least
quadratically at a point, say, $x_1$, then, it is clear that
the resulting potential $V(x)$, in addition to satisfying the
above mentioned conditions (i.e. having a maxima at the origin
and a minima at infinity), will also contain a point of
inflection at $x_1$.
A simple function that satisfies our requirements turns out to
be\footnote{Actually, this function contains another point of
inflection at the origin. But, as we shall restrict ourselves
to the domain $x>0$,  it is not useful to us.}
\beq
f(x)=\l[(x-x_1)^2\; x^2\r].
\eeq
For this choice of the function $f(x)$ and appropriate values
of the parameters $\lambda$ and $T_{0}$, we find that the
corresponding potential gives rise to punctuated inflation.
However, it is important to note that, unlike the earlier
examples, the rapid roll phase {\it does not}\/ result in a
deviation from inflation.
In Fig.~\ref{fig:ti}, we have plotted the scalar and the tensor
power spectra, and the corresponding tensor-to-scalar ratio
that we obtain in this case.
It is clear from the figure that the spectra broadly behave in
the same fashion as in the earlier examples.


\section{The effects on the B-modes of the CMB}\label{sec:clbb}

As is well known, the polarization of the CMB can be decomposed
into the E and B-components.
While the E-mode polarization is affected by both the scalar as well
as the tensor perturbations, the B-modes are generated {\it only}\/
by the tensor perturbations\footnote{In fact, the B-modes are
created by the vector perturbations too.
However, inflation does not generate any vector perturbations.}.\/
Therefore, the B-mode provides a direct signature of the
primordial tensor perturbations (see, for instance, the last
reference in Ref.~\cite{texts}).
The detection of the B-mode is a coveted, prime goal of the
experimental community (see, for example, the white
paper~\cite{cmbp-dwp}).
We feel that the punctuated inflationary scenario can provide
additional theoretical motivation for this endeavor.

\begin{figure}[!htb]
\begin{center}
\vskip 25pt
\resizebox{270pt}{180pt}{\includegraphics{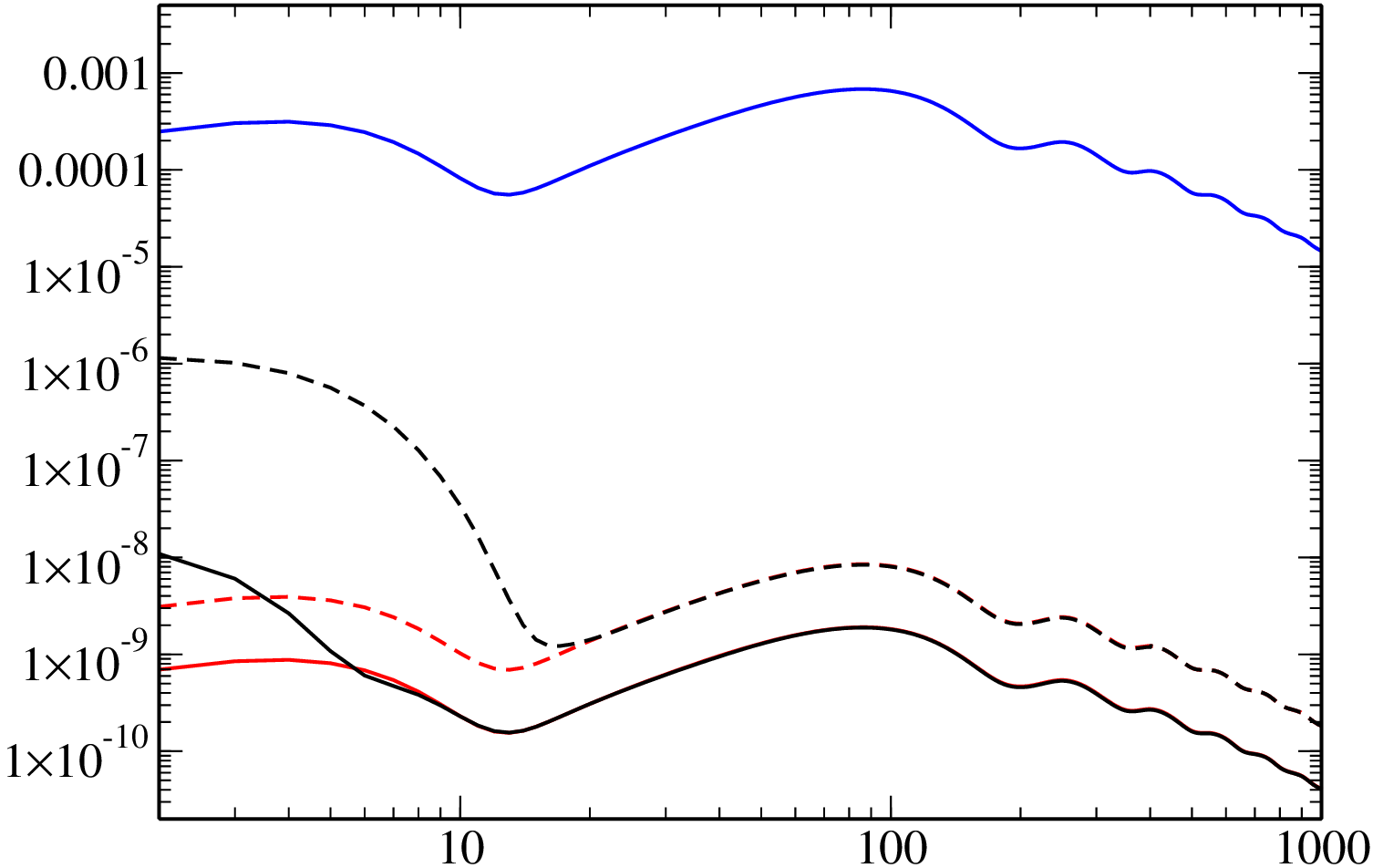}}
\vskip -170 true pt \hskip -295 true pt
\rotatebox{90}{$\l[\ell\, (\ell+1)\;
C_{\ell}^{\rm BB}/(2\,\pi)\r]\; \mu\, {\rm K}^{2}$}
\vskip 65 true pt \hskip 10 true pt $\ell$
\caption{The B-mode CMB angular power spectrum $C_{\ell}^{\rm BB}$
has been plotted as a function of the multipole~$\ell$ for the best
fit values of the $n=3$ (the solid black line) and the $n=4$ (the 
dashed black line) cases of the MSSM-motivated model.
For comparison, we have also plotted the $C_{\ell}^{\rm BB}$ for the
concordant cosmological model with a strictly scale invariant tensor
spectrum and a tensor-to-scalar ratio of $r=0.01$ (the solid blue line),
$r=2\times10^{-8}$ (the solid red line) and $r=10^{-7}$ (the dashed
red line).
The latter two curves match the $n=3$ and $n=4$ cases of the MSSM-motivated 
model at the small angular scales, and they help in highlighting the 
effects of punctuated inflation at the lower multipoles.}
\label{fig:clbb}
\end{center}
\end{figure}

We have evaluated the angular power spectrum of the B-mode
polarization of the CMB (i.e. $C_{\ell}^{\rm BB}$) using the
Boltzmann code CAMB~\cite{camb}.
In Fig.~\ref{fig:clbb}, we have plotted $C_{\ell}^{\rm BB}$ for
the best fit values of the parameters in the $n=3$ and the $n=4$
cases of the MSSM-motivated model.
For comparison, we have also plotted the corresponding angular power
spectra for the concordant cosmological model with a strictly scale
invariant tensor spectrum and a tensor-to-scalar ratio of $r=0.01$,
$r=2\times10^{-8}$ and $r=10^{-7}$ (the last two values have been
chosen since they match the $n=3$ and $n=4$ cases of the 
MSSM-motivated model at the small angular scales).
The $C_{\ell}^{\rm BB}$ for the two cases of the MSSM-motivated model 
clearly exhibit an
increase in their amplitude at the lower multipoles, reflecting
the rise in the tensor-to-scalar ratio on these scales\footnote{We
should point out that, in order to evaluate the CMB angular power
spectra, CAMB integrates over the following range of wavenumber of
the primordial scalar and tensor spectra: $7.5\times 10^{-6}\lesssim
k\lesssim 2.8 \times 10^{-1}\; {\rm Mpc}^{-1}$.
Note that, in both the MSSM-motivated cases, the region where $r>1$ is well
beyond the Hubble scale today, viz. $k \simeq 10^{-4}\; {\rm Mpc}^{-1}$
(cf.~Fig.~\ref{fig:r}).
Therefore, the resulting $C_{\ell}^{\rm BB}$ will be sensitive to the
tensor power at such large scales, and we need to be careful about the
lower limit of the $k$ integral in CAMB.
We find that the CAMB's default lower limit works well in these cases
since $r$ attains its maximum value at a wavenumber that is larger than
the lower limit.}.
But, despite the rise at the lower multipoles, the amplitude of
$C_{\ell}^{\rm BB}$ in these cases proves to be way too
smaller than what is possibly detectable in the near future
(current/upcoming missions such as PLANCK~\cite{planck} and
CMBPol~\cite{CMBPol} are expected to be sensitive to
$r \gtrsim 0.01$).
However, since the increase in the B-mode power at large
angular scales is a generic feature of punctuated inflation,
we feel that it improves the possibility that the effect may
be detected in the future.
It is conceivable that there exist punctuated inflationary models
that predict a significantly larger tensor-to-scalar ratio,
while still providing a good fit to the CMB temperature angular
power spectrum.
For example, given a value of the tensor-to-scalar ratio at, say,
the Hubble radius today, one can possibly work in the slow roll limit
and invert the scalar power spectrum (that results in a good fit) to
arrive at a suitable inflationary potential.
It seems a worthwhile exercise to systematically hunt for such models.


\section{Conclusions}\label{sec:concl}

In our earlier work~\cite{pi}, we had performed a Markov Chain Monte
Carlo analysis to determine the values of the parameters of the 
MSSM-motivated model that provide the best fit to the WMAP $5$-year 
data for the CMB angular power spectrum.
We had found that a scalar spectrum in the $n=3$ case leads to a much
better fit of the observed data than the spatially flat, $\Lambda$CDM
model with a power law, primordial spectrum.
We should emphasize again that we have not carried out such a comparison
with the data for the hybrid or the tachyon model.
In the $n=3$, MSSM-motivated model, we had found that, in addition to the 
drop in the power at large scales, the bump present in the spectrum before
it turns nearly scale invariant had led to the improvement in the fit.
In the $n=4$ case, a rather large bump had led to a poor fit to the data.
We find that, a similar, large bump arises in the hybrid model as well.
Also, as we had mentioned, in the hybrid model, the scalar spectral index
proves to be greater than unity at small scales.
We feel that these two features will not allow a better fit in the
hybrid case.
In the tachyonic model, though the spectral index is close to the
observed value, we find that no bump ({\it above}\/ the asymptotic,
nearly scale invariant amplitude) arises in the spectrum.
We expect that this feature will spoil the fit to the data.
Moreover, we believe that the lack of such a bump is due to the fact
that inflation is not interrupted in this case.

In models which start with a period of fast roll, along with
the scalar power, the tensor power is also suppressed at large
scales~\cite{nicholson-2008}.
But, the drop in the scalar power proves to be sharper than that
of the tensors and, as a result, the tensor-to-scalar ratio
displays a rise over these scales in such models.
It has been argued that such a feature may be detected by
ongoing missions such as, for instance, PLANCK~\cite{planck}.
We too encounter an increase in the tensor-to-scalar ratio on
the large scales, though the reason is somewhat different.
In punctuated inflation, the rise in the tensor-to-scalar ratio
turns out to be much stronger due to the fact that the tensor
amplitude itself increases on large scales.
Intriguingly, we find that the rapid rise leads to the
tensor-to-scalar ratio being much larger than unity for a small
range of modes.
However, in the specific models we have considered, the tensor
amplitude on scales of cosmological interest (say, $10^{-4} <
k < 1\; {\rm Mpc}^{-1}$) proves to be too small ($r <10^{-4}$)
for the effect to be possibly detected in the very near future.

The sharper the drop in the scalar spectrum at large scales, the
better seems to be the fit to the low CMB quadrupole.
In punctuated inflation, the steeper the drop in the scalar power,
the faster will be the corresponding rise in the tensor power at
large scales.
Therefore, if the scalar power drops fast, the tensor-to-scalar ratio
can be larger on the small scales, thereby improving the prospects of
its detection through the B-modes of the CMB polarization.
However, empirical evidence indicates that, in punctuated inflation,
a steeper drop in the scalar power requires a larger value of the
first slow roll parameter $\epsilon$ during the rapid roll.
But, such a large $\epsilon$ also leads to a bigger bump (above the
asymptotic amplitude) in the scalar spectrum before it turns scale
invariant.
While a suitable bump seems to provide a better fit to the data at
a few lower multipoles after the quadrupole, too large a bump seems
to spoil the fit to the data (as in the $n=4$, MSSM-motivated model).
In other words, to lead to a good fit, there appears to be a trade
off between the sharpness of the cut off and the size of the bump
in the scalar power spectrum.
We are currently exploring punctuated inflationary models that will
lead to a sufficiently steep drop in the scalar power at large scales,
a suitably sized bump at the top of the spectrum, and also a reasonable
tensor amplitude at small scales that may be detectable by forthcoming
missions.


\section*{Acknowledgments}

We would like to thank Jinn-Ouk Gong for discussions during the
early stages of this work, and for his comments on the manuscript.
TS acknowledges enlightening discussions with Alexei Starobinsky.
We would also like to acknowledge the use of the high performance 
computing facilities at the Harish-Chandra Research Institute, 
Allahabad, India, and at the Korea Institute for Advanced Study, 
Seoul, Korea.


\appendix

\section{The evolution of the scalar and tensor perturbations
for a mode with $r>1$}\label{sec:app}

\begin{figure}[!htb]
\begin{center}
\vskip 25pt
\resizebox{270pt}{180pt}{\includegraphics{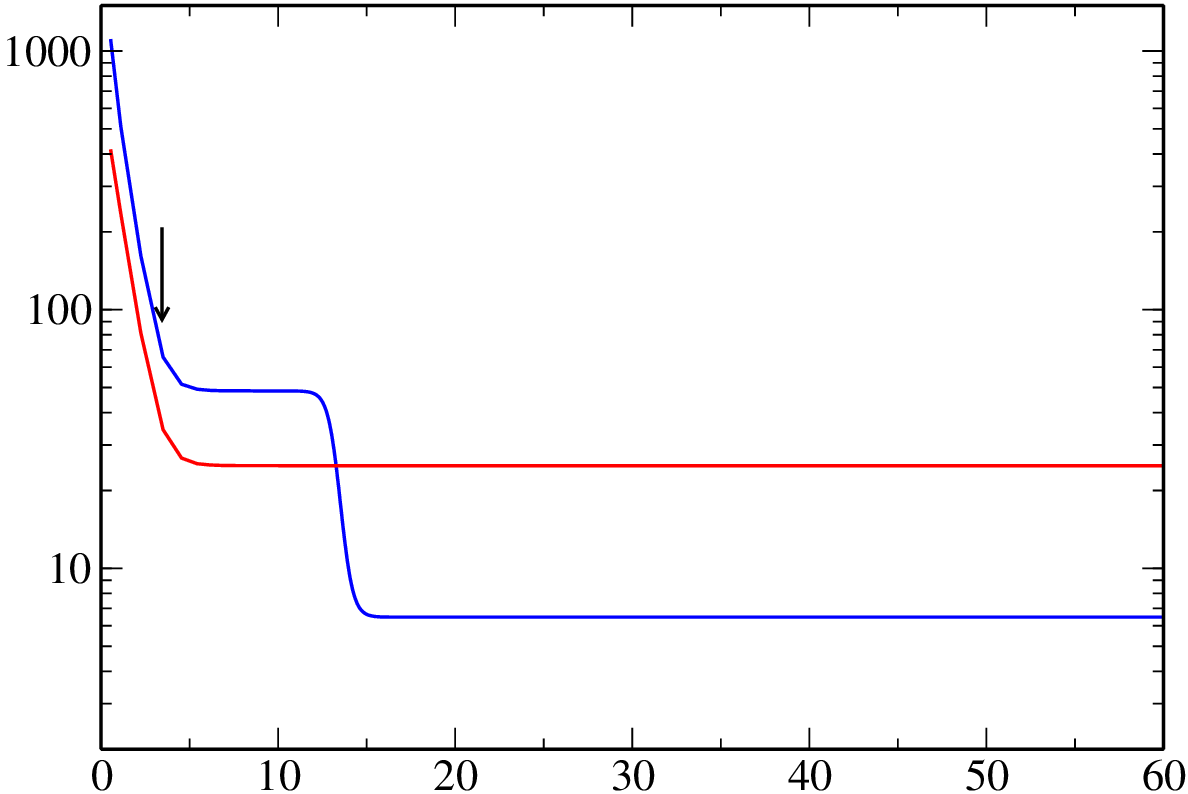}}
\vskip -155 true pt \hskip -310 true pt
{$\vert{\cal R}_{\rm k}\vert$, $\vert{\cal U}_{\rm k}\vert$}
\vskip 155 true pt \hskip 16 true pt $N$
\caption{The evolution of the amplitudes of the curvature
perturbation ${\cal R}_{\rm k}$ (in blue) and the tensor
perturbation ${\cal U}_{\rm k}$ (in red) has been plotted
as a function of the number of $e$-folds~$N$ for the best
fit values of the $n=3$, MSSM-motivated model.
These perturbations correspond to the mode $k=10^{-5}\;
{\rm Mpc}^{-1}$, and the arrow denotes the time when the
mode leaves the Hubble radius.
Notice that, as expected, the tensor amplitude freezes at its
value near Hubble exit.
In contrast, the amplitude of the curvature perturbation is
suppressed on super-Hubble scales.
Evidently, it is this behavior of the curvature perturbation
that leads to the large tensor-to-scalar ratio associated with
the mode.}
\label{fig:st}
\end{center}
\end{figure}

In the various examples of punctuated inflation that we had
discussed in the text, though the tensor-to-scalar ratio
remains too small ($r <10^{-4}$) on the scales of cosmological
interest, we find that there exists a small range of modes for
which the tensor-to-scalar ratio turns out to be greater than
unity.
We believe that this is an interesting feature with potentially
observable consequences.
To highlight this feature, in Fig.~\ref{fig:st}, we have plotted
the evolution of the amplitudes of the curvature and the tensor
perturbations for a mode that has a tensor-to-scalar ratio greater
than unity in the $n=3$, MSSM-motivated model.
Note that, due to the deviation from slow roll, on super-Hubble
scales, the amplitude of the curvature perturbation is suppressed
when compared to its value near Hubble exit.
Whereas, the tensor amplitude approaches a constant value soon
after the mode leaves the Hubble radius.
It is such a suppression of the curvature perturbation that results
in the tensor-to-scalar ratio being greater than unity.


\end{document}